\documentclass[aps,amsmath,pra,reprint,groupedaddress]{revtex4-1}
\usepackage{graphicx}
\usepackage{dcolumn}
\usepackage{bm}
\usepackage{amsfonts}
\usepackage{array}
\usepackage[colorlinks,linkcolor=blue,anchorcolor=blue,citecolor=blue]{hyperref}

\begin{document}

\title{Finite-size analysis of continuous variable source-independent quantum random number generation}
\author{Junyu Zhang\textsuperscript{1} }
\author{Yichen Zhang\textsuperscript{1}}
    \email[Correspondence: ]{zhangyc@bupt.edu.cn.}
\author{Ziyong Zheng\textsuperscript{1}}
\author{Ziyang Chen\textsuperscript{2}}
\author{Bingjie Xu\textsuperscript{3}}
\author{Song Yu\textsuperscript{1}}
\affiliation{\textsuperscript{1}State Key Laboratory of Information Photonics and Optical Communications, Beijing University of Posts and  Telecommunications, Beijing, 100876, China }
\affiliation{\textsuperscript{2}State Key Laboratory of Advanced Optical Communication Systems and Networks, School of Electronics Engineering and Computer Science, and Center for Quantum Information Technology, Peking University, Beijing 100871, China }
\affiliation{\textsuperscript{3}Science and Technology on Security Communication Laboratory, Institute of Southwestern Communication, Chengdu 610041, China}

\date{\today} 

\begin{abstract}
We study the impact of finite-size effect on continuous variable source-independent quantum random number generation. The central-limit theorem and maximum likelihood estimation theorem are used to derive the formula which could output the statistical fluctuations and determine upper bound of parameters of practical quantum random number generation. With these results, we can see the check data length and confidence probability has intense relevance to the final randomness, which can be adjusted according to the demand in implementation. Besides, other key parameters, such as sampling range size and sampling resolution, have also been considered in detail. It is found that the distribution of quantified output related with sampling range size has significant effects on the loss of final randomness due to finite-size effect. The overall results indicate that the finite-size effect should be taken into consideration for implementing the continuous variable source-independent quantum random number generation in practical.
\end{abstract}
\maketitle

\section{Introduction} \label{sec1}
Random numbers are traditionally generated using deterministic processes. However, these predictable random numbers will lead to security loopholes or errors in some fields, such as cryptography, scientific simulation and lottery. As an ideal alternative, quantum random number generators (QRNGs) have attracted much attention in the past few years~\cite{herrero2017quantum,ma2016quantum} because they exploit the fundamental indeterminacy of quantum mechanics and can generate unpredictable random numbers. Based on the randomness reliability levels, three types can be categorized in QRNGs: practical QRNGs, device-independent QRNGs and semi-device-independent QRNGs. Practical QRNGs~\cite{nguyen2018programmable,wang2015bias,yan2015high,Stefanov00,wayne2018post,Ren11,zheng20196,gehring20188,shakhovoy2019quantum,Wei09} completely trust their well characterized devices and generate random numbers with high speed. Device-independent(SI) QRNGs~\cite{pironio2010random,liu2018high} have the most
paranoid assumption that the devices are all untursted and have highest security levels, which results to low generation speed that can hardly meet actual demands. As a trade off, semi-device-independent QRNGs generate high-speed random numbers by just trusting parts of the devices. One mainly considers QRNGs with untrusted measurement devices~\cite{Nie16,Cao15} or untrusted sources in this filed. Therein, source-independent (SI) QRNGs get people's favour due to the difficulty of pure source state preparation~\cite{cao2016source,smith2019simple,marangon2017source,xu2019high,avesani2018source}.

Imparity with low-speed discrete variable SI-QRNGs~\cite{vallone2014quantum}, continuous variable (CV) SI-QRNGs achieve higher speed. CV-SI-QRNGs eliminate the effects of malicious eavesdropper through randomly switching the measurement bases generally with phase modulator. The application of phase-randomized local oscillator with gain-switched laser makes it easier to implement~\cite{smith2019simple}. The heterodyne measurement remarkably increases the generating speed to $17$ Gbps and operates without an initial source of randomness~\cite{avesani2018source}. Several practical imperfections have been considered in~\cite{xu2019high} and achieve  $15.07$ Gbps high generation speed. In most CV-SI-QRNG protocols, the mutual information between Alice and Eve is the most crucial parameter, which should be estimated with generated raw sequences in practise. To guarantee the security, CV-SI-QRNGs always discard untrusted parts of information by post-processing algorithms, such as Toeplitz hashing matrix, after parameter estimation.

The estimation of precise parameters requires infinite-size of check data. However, the size of check data in practise is finite, because any practical system can only run in finite time. This will lead to the statistical fluctuations of estimated parameters, i.~e.~we always overestimate or underestimate the parameters compared with their ideal values. This may result in miscalculating the available randomness and lead to security loopholes. Thus, the finite-size effect is necessary to be considered in practical systems.

In this paper, we study finite-size effect on the CV-SI-QRNG protocol~\cite{xu2019high} by calculating the statistical fluctuations of estimated parameters. The central limit theorem and maximum likelihood estimation theorem are utilized to find the distribution of estimated parameters. The statistical fluctuation of final randomness can be computed out with a given confidence probability. Then, we give numerical simulations on extractable randomness with different variable parameters including check data length $m$, confidence probability $\epsilon$, sampling range size $N$ and sampling resolution $n$ for studying the impacts of finite-size effect under various scenarios. Finally, we analyze the relationship between sampling range size and finite-size effect in detail.

In this paper, we introduce our work as follows: In Sec.~\ref{sec2} we review the CV-SI-QRNG protocol proposed in ~\cite{xu2019high}. In Sec.~\ref{sec3} we briefly describe the security analysis of our protocol, then consider the effect of finite size on the security of the protocol in detail and derive the formulas for the analysis of finite-size. In Sec.~\ref{sec4} we show and discuss the results of simulation with several variable parameters which could impact the finite-size effect. At last, we give conclusions in Sec.~\ref{sec5}.
\section{Continuous variable source-independent QRNG protocol} \label{sec2}
\begin{figure}
\centering
\includegraphics[width=.5\textwidth]{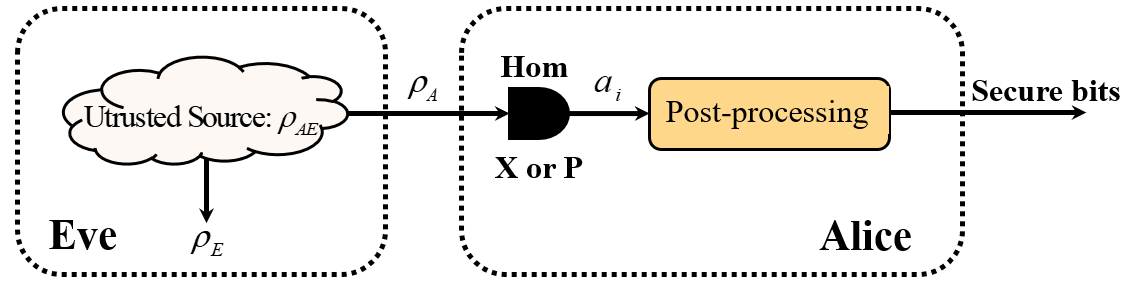}
\caption{(Color online) Schematic of CV-SI-QRNG protocol. Eve completely controls the untrusted source and keeps quantum state $\rho_E$ related with quantum state $\rho_A$ which is sent to Alice. Then Alice does balanced homodyne measurement on $\rho_A$ with bases which switched randomly between $X$ and $P$ quadratures. The results of measurement $a_i$ must be eliminated the insecurity parts by post-processing to output available security random numbers.}
\label{img1}
\end{figure}

The more detailed introduction about the CV-SI-QRNG protocol is in~\cite{xu2019high}, we just make a brief introduction here. Fig.~\ref{img1} illustrates the schematic of our CV-SI-QRNG protocol. There is an untrusted and uncharacterized source prepared by Eve with quantum state $\rho _A$, which is allowed in infinite dimensions. Eve keeps a quantum system $E$ related to system $A$, and then sends $\rho _A$ to Alice for homodyne measurement. To extract security randomness, Alice switches the measurement bases randomly between $X$ and $P$ quadratures for measuring $\rho _A$ with an initial random seed.

Obviously, the switch should be in an independently and identically distribution (i.~i.~d.) way to extract i.~i.~d. random numbers. Now we define $n_{tot}$ as the number of total measurement result, $n_{c}$ as the number of check samples out of total measurements and $t$ as the length of random seed. It should be noted that we make an assumption that the measurement can be trusted and calibrated well, meanwhile a quantum correlated Eve can introduce all the excess noise.

After Alice measures the $X$ and $P$ quadratures and gets check sequence, the covariance matrix (CM) of the state $\rho _A$ can be estimated with this formula:
\begin{equation}
{\gamma _A} = \left[ {\begin{array}{*{20}{c}}
{{V_x}}&c\\
c&{{V_p}}
\end{array}} \right],\label{equ1}
\end{equation}
where $V_x$ and $V_p$ are the variance of  $X$ and $P$ quadratures, and $c$ is the co-variance between $X$ and $P$ quadratures for $\rho _A$.

At last, the final randomness that Alice could extract from total $n_{tot}$ measurement is asymptotically calculated by $\left( {{n_{tot}} - {n_c}} \right)\left( {H\left( {{a_i}} \right) - S\left( {{a_i}:E} \right)} \right) - t$, where $H(a_i)$ is the discrete Shannon entropy of Alice's measurement results and $S(a_i:E)$ is the quantum mutual information between the quantum state $\rho_E$ kept by Eve and Alice's measurement results. Then putting the remaining measurement results into randomness extractor. There is several mature and verifiable  randomness extractors that can eliminate the untrusted part of randomness and remain secure true randomness based on the parameters set by Alice.

In the whole protocol we introduced above, no assumptions have been made on the source of QRNG. That means, the random numbers extracted are secure enough regardless of the purity and dimensions of Alice's input state. The measurement devices have our fully trust since it's inaccessible for Eve.

\section{security analysis and finite-size effect }
\label{sec3}
In this section
, we will make a brief analysis about the security of our CV-SI-QRNG protocol firstly (the more detailed analysis is in~\cite{xu2019high}) and the formulas that are used in finite-size analysis and numerical simulations are derived.

\subsection{Security analysis}
\label{security analysis}
Our protocol is very similar with continuous-variable quantum key distribution~\cite{leverrier2010finite,jouguet2012analysis,zhang2017finite}, the key rate of our CV-SI-QRNG is calculated with the Devetak-Winter formula~\cite{devetak2005distillation}, in the case of i.~i.~d, which is used in continuous-variable quantum key distribution, is given by:
\begin{equation}
K = \beta \left( {I\left( {a:b} \right) - S\left( {a:E} \right)} \right),\label{equ2}
\end{equation}
where $\beta$ is the reconciliation efficiency, $I(a:b)$ is the classical mutual information between Alice and Bob and $S(a:E)$ is the Holevo bound between the quantum state $\rho _E$ kept by Eve and the result $a$ measured by Alice. Different from continuous variable quantum key distribution scenario, in QRNG, Alice and Bob are at the same station. Thus, we don't perform information reconciliation. Therefore the mutual information $I(a:b)$ is described by Shannon entropy $H(a)$ which is just defined by Alice's actually obtained data and the reconciliation efficiency $\beta$ equals to 1.

In the Eq.~\ref{equ2}, it should be noted that Alice's measurement result $a$ is a continuous variable, that means the homodyne detection is perfect with infinite range and resolution. But, in practical, the measurement of Alice must be imperfect with finite range and finite resolution, i.~e.~it should be coarse-gained and discrete~\cite{zhang20181}. It mainly depends on the precision of analog-to-digital convertor, which digitizes the unreadable continuous data $a$ into $n$ bit $a_i$. We define the sampling range of the analog-to-digital convertor is $\left[ {{\rm{ - N + }}\Delta {\rm{/2,N - }}\Delta {\rm{3/2}}} \right]$, where $\Delta {\rm{ = }}N/{2^{n - 1}}$. The sampling range has been divided evenly into ${2^{\rm{n}}}$ bins and $\Delta$ is the interval between every adjacent $a_i$.

Then, the available secure randomness that uncorrelated with state $\rho _E$ kept by Eve can be calculated out in our condition by formula:
 \begin{equation}
{R_{dis}}\left( {{a_i}\left| E \right.} \right) = H\left( {{a_i}} \right) - S\left( {{a_i}:E} \right),\label{equ3}
\end{equation}
where $H\left( {{a_i}} \right)$ is the discrete Shannon entropy of $a_i$, $S\left( {{a_i}:E} \right)$ is the Holevo bound between the quantum state $\rho _E$ kept by Eve and the discrete result $a_i$ measured by Alice. $H\left( {{a_i}} \right)$ can be calculated easily with Alice's measurement result $a_i$. So, we focus on $S\left( {{a_i}:E} \right)$ to find its upper bound due to the difficulty of calculating it out directly.

First, there is an inequality proved in \cite{xu2019high} that $S\left( {{a_i}:E} \right) \le S\left( {a:E} \right)$. It means the digitization operation won't increase the mutual information between the quantum state $\rho_E$ kept by Eve and the discrete result $a_i$ measured by Alice. Then, we consider the worst case that the state $\rho_E$ kept by Eve is a purification of $\rho_A$. This leads to $S\left( {{\rho _E}\left| a \right.} \right)=0$ and $S(a:E) = S({\rho _E}) = S\left( {{\rho _A}} \right)$. The von Neumann entropy of Gaussian state $\rho ^G$ is the upper bound in that of arbitrary state $\rho$ if they have the same CM~\cite{wolf2006extremality}. So, there is $S\left( {{\rho _A}} \right) \le S\left( {\rho _A^G} \right)$.

Now, the lower bound of final extractable randomness~\cite{ma2013postprocessing} goes to
\begin{equation}
{R_{dis}}\left( {{a_i}\left| E \right.} \right) \ge H\left( {{a_i}} \right) - S\left( {\rho _A^G} \right).\label{equ4}
\end{equation}

Focusing on $S\left( {\rho _A^G} \right)$, its von Neumann entropy is
\begin{equation}
S\left( {\rho _A^G} \right) = \frac{{\lambda  + 1}}{2}{\log _2}\frac{{\lambda  + 1}}{2} - \frac{{\lambda  - 1}}{2}{\log _2}\frac{{\lambda  - 1}}{2},\label{equ5}
\end{equation}
where $\lambda$ is the symplectic eigenvalue of matrix in Eq.\ref{equ1} and $\lambda {\rm{ = }}\sqrt {\det ({\gamma _A})}  = \sqrt {{V_x}{V_p} - {c^2}}$. But we cannot obtain the exact value of $\lambda$ due to the loss information inside the digitized interval. Therefor, we have to estimate its upper bound to obtain secure randomness. We set $c=0$ and defined $\lambda  < \bar \lambda  = \sqrt {{{\bar V}_x}{{\bar V}_p}}$, where $\bar V_x$ is the upper bound of $V_x$, given by:
\begin{equation}
\begin{array}{c}
{{\bar V}_x} = {p_{dis}}\left( {{a_{{i_{\min }}}}} \right){\left( {{a_{\min }} - \bar a} \right)^2} + {p_{dis}}\left( {{a_{{i_{\max }}}}} \right){\left( {{a_{\max }} - \bar a} \right)^2}\\
 + {\sum\limits_{i = {i_{\min }} + 1}^0 {{p_{dis}}\left( {{a_i}} \right)\left( {{a_i} - \bar a - \frac{1}{2}\Delta } \right)} ^2}\\
 + {\sum\limits_{i = 1}^{{i_{\max }} - 1} {{p_{dis}}\left( {{a_i}} \right)\left( {{a_i} - \bar a + \frac{1}{2}\Delta } \right)} ^2},
\end{array}\label{equ6}
 \end{equation}
where ${p_{dis}}\left( {{a_i}} \right)$ is the probability of $a_i$, $\bar a$ is the mean of Alice's measurement result, $a_{min}$  and $a_{max}$ are the minimal and maximum values of a large bound $\left[ {{\rm{ - }}{a_{\lim }},{a_{\lim }}} \right]$. This bound is set to make the probability that measurement result $a$ is outside the bound is negligible for more rigorous proof. As shown in Eq.~\ref{equ6}, we use the boundary of the discrete interval which is farther from 0 point to estimate the upper bound of $V_x$, i.~e.~we use ${a_i} + \frac{1}{2}\Delta$ if ${a_i} > 0$ and use ${a_i} - \frac{1}{2}\Delta$ if ${a_i} < 0$. We give the same definitions and assumptions for $\bar V_p$.

Our protocol uses vacuum state as randomness source which is not be disturbed by Eve. And we define the variance of vacuum fluctuation as $\sigma _{{\rm{vac}}}^2{\rm{ = 1}}$ for normalized calculation. In practice, we must consider the excess noise $\varepsilon$ which is mixed in quantum state $\rho_A$ due to the side information. Thus, the  expected variance of Alice's measurement result becomes ${\sigma ^2} = {V_x} = {V_p} = \varepsilon {\rm{ + }}\sigma _{{\rm{vac}}}^2$. Alice's homodyne measurement result $a$ is a continuous variable which satisfies Gaussian distribution with variance $\sigma^2$ and zero mean. The  expected discrete probability ${p_{dis}}\left( {{a_i}} \right)$ of $a_i$ can be calculated out based on the Gaussian distribution.

No assumptions about the source have been made in the security analysis. Therefore the final extracted randomness is source-device-independent and does not relate to Eve.

\subsection{Finite-size effect}
\label{finite-size effect}
In fact, our CV-SI-QRNG and data sampling can't run forever. Thus, we could only extract finite-size sequences for random numbers. Some parameters of the extracted data will have fluctuations and deviations compared with their ideal values. In our protocol, the number of total measurement results $n_{tot}$ and the number of check samples $n_c$ are finite absolutely. Therefore the extractable randomness in Eq.~\ref{equ4} should be bounded with finite-size effect.

Our analysis of finite-size effect focuses on $\bar V_x$ which is defined in Eq.~\ref{equ6} in infinite-size scenario. The formula we used for obtaining $V_x$ with discrete sequence $a = \{ {a_1},{a_2} \cdots {a_m}\} $ which is the output of convertor is
\begin{equation}
{\hat V_x} = \frac{1}{m}{\sum\limits_{i = 1}^m {\left( {{a_i} \pm \frac{1}{2}\Delta -\bar a} \right)} ^2},\label{equ7}
\end{equation}
where $m$ is the check data length and $\hat V_x$ is the true value of $V_x$. For the plus or minus sign in front of $\frac{1}{2}\Delta$, we take plus sign if $a_i>0$, and take minus sign if $a_i<0$. Now, we expand Eq.~\ref{equ7} to
\begin{equation}
\begin{array}{c}
{{\hat V}_x} = \frac{1}{m}\sum\limits_{i = 1}^m {\left( {a_i^2 \pm \Delta  \cdot {a_i}} \right)}  + \frac{1}{m}\sum\limits_{i = 1}^m {\left( { \mp \Delta  \cdot \bar a} \right)} \\
 + \frac{1}{m}\sum\limits_{i = 1}^m {\left( {{{\bar a}^2} - 2{a_i}\bar a} \right)}  + \frac{1}{4}{\Delta ^2}.
\end{array}\label{equ8}
\end{equation}

An approximation could be done that $\pm \Delta  \cdot {a_i} = \left| {\Delta  \cdot {a_i}} \right|$ based on the selection of plus or minus sign. For a large $m$, the numbers of plus sign and minus sign we take in Eq.~\ref{equ7} are unlikely to differ by orders of magnitude, and $\bar a$ is almost close to 0. Therefore, we can make an approximation that $\frac{1}{m}\sum\nolimits_{i = 1}^m {\left( { \mp \Delta  \cdot \bar a} \right)}  \approx 0$. Take these results into Eq.~\ref{equ8}, we get
\begin{equation}
{\hat V_x} = \frac{1}{m}\sum\limits_{i = 1}^m {\left( {a_i^2 + \left| {\Delta  \cdot {a_i}} \right|} \right)}  - {\bar a^2} + \frac{1}{4}{\Delta ^2}.\label{equ9}
\end{equation}

\begin{figure}[b]
  \centering
  \includegraphics[width=.5\textwidth]{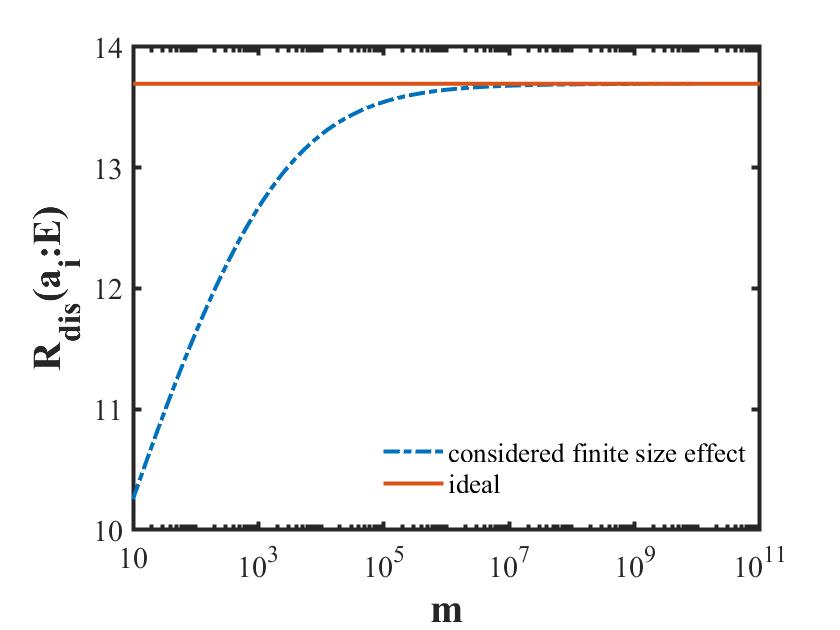}
  \caption{(Color online) The impact of check data length $m$ for finite-size effect on extractable randomness $R_{dis}(a_i:E)$. We set the excess noise $\varepsilon=0.1$, ideal variance $\sigma^2=1+\varepsilon=1.1$, confidence probability $\epsilon=10^{-10}$, sampling resolution $n=16$, and sampling range size $N=3\sigma$ here.}
  \label{img2}
\end{figure}

It should be taken note that $\bar a$ has correlation with $a_i$, but for a large $m$ the correlation is very small, and $\bar a$ is close to 0 compared with $a_i^2$. So we use ideal value ${\mu _a}$ instead of $\bar a$. We define $a_i^2 + \left| {\Delta  \cdot {a_i}} \right| = {b_i}$ and get
\begin{equation}
{\hat V_x} = \frac{1}{m}\sum\limits_{i = 1}^m {{b_i}}  - {\mu _a^2} + \frac{1}{4}{\Delta ^2}.\label{equ10}
\end{equation}

Let's focus on $\sum\nolimits_{i = 1}^m {{b_i}}$. In Sec~\ref{sec2}, we have assumed that the experiments are repeated in i.~i.~d. way, thus the extracted numbers $a_i$ are i.~i.~d. too. For this reason, $b_i$ is i.~i.~d. absolutely based on its definition. According to central limit theorem~\cite{rosenblatt1956central}: for an i.~i.~d. random number sequence $b = \{ {b_1},{b_2} \cdots {b_m}\} $, if $m$ is large enough (more than $10^4$), it has following approximation
\begin{equation}
\frac{{\sum\nolimits_{i = 1}^m {{b_i} - m{\mu _b}} }}{{\sqrt m {\sigma _b}}} \sim N\left( {0,1} \right),\label{equ11}
\end{equation}
where $\mu_b$ and $\sigma_b$ are the ideal mean value and standard variance of $b$ respectively and $N(0,1)$ is standard Gaussian distribution. Thus the value of $\frac{1}{m}\sum\nolimits_{i = 1}^m {{b_i}}$ satisfies Gaussian distribution approximately:
\begin{equation}
\frac{1}{m}\sum\limits_{i = 1}^m {{b_i}}  \sim N\left( {{\mu _b},\frac{{\sigma _b^2}}{m}} \right),\label{equ12}
\end{equation}
and $\hat V_x$ satisfies
\begin{equation}
{\hat V_x} \sim N\left( {{\mu _b} - \mu _a^2 + \frac{1}{4}{\Delta ^2},\frac{{\sigma _b^2}}{m}} \right).\label{equ13}
\end{equation}

\begin{figure}[b]
  \centering
  \includegraphics[width=.5\textwidth]{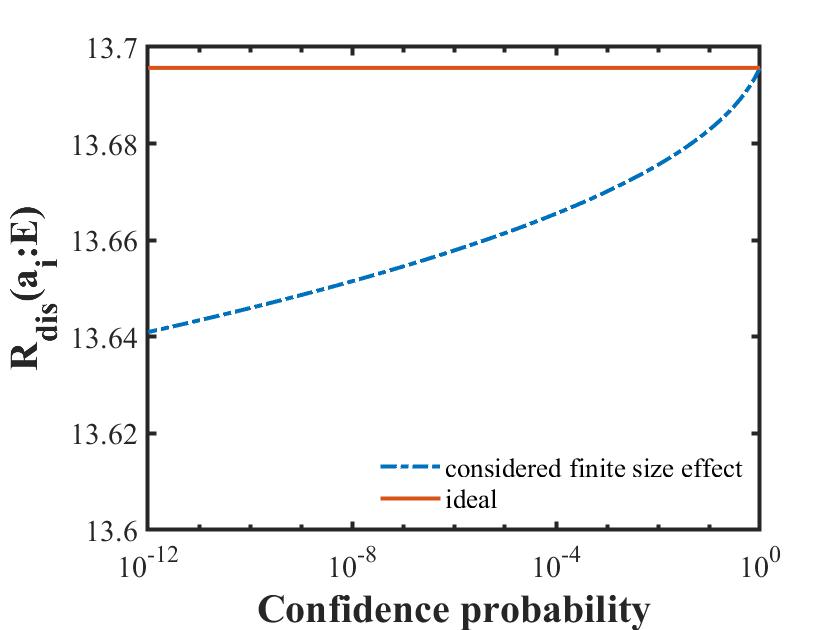}
  \caption{(Color online) The impact of confidence probability $\epsilon$ for finite-size effect on extractable randomness $R_{dis}(a_i:E)$. We set the excess noise $\varepsilon=0.1$, ideal variance $\sigma^2=1+\varepsilon=1.1$, check data length $m=10^6$, sampling resolution $n=16$, and sampling range size $N=3\sigma$ here.}
  \label{img3}
\end{figure}

In Eq.~\ref{equ6}, an assumption has been made that the minimal and maximum values of $a$ are determined by a large bound $\left[ {{\rm{ - }}{a_{\lim }},{a_{\lim }}} \right]$. So, we make the same assumption here for calibrating $\mu_b$, $\sigma_b$ and $\mu_a$. Then we find ${\mu _b} - \mu _a^2 + \frac{1}{4}{\Delta ^2}{\rm{ \approx }}\bar V$. It means our method estimates $\hat V_x$ quite well.

Now we can estimate the confidence interval of $\hat{V_x}$ with confidence probability $\epsilon$:
\begin{equation}
{\hat V_x} \in \left[ {{{\bar V}_x} - \Delta V,{{\bar V}_x} + \Delta V} \right],\label{equ14}
\end{equation}
where
\begin{equation}
\Delta V = {Z_{\epsilon/2}}\sqrt {\frac{{\sigma _b^2}}{m}},\label{equ15}
\end{equation}
and $Z_{\epsilon/2}$ satisfies $1 - erf\left( {{Z_{\epsilon/2}}/\sqrt 2 } \right) = \epsilon$. $\epsilon$ is the probability of the $\hat V_x$ estimation outside the confidence interval. And ${erf}$ is the error function defined as
\begin{equation}
{erf\left( x \right) = \frac{2}{{\sqrt \pi  }}\int_0^x {{e^{ - {t^2}}}dt}}.\label{equ16}
\end{equation}
To get the upper bound of $\lambda$ in Eq.~\ref{equ5}, we use $\hat V_x$ to estimate $V_x$. So there is
\begin{equation}
{V_{x\max }} \approx {\bar V_x} + {Z_{\epsilon/2}}\sqrt {\frac{{\sigma _b^2}}{m}}.\label{equ17}
\end{equation}

Similarly, we consider finite-size effect on $V_p$ with the same method, and get $V_{p\max}$. Putting these parameters into Eq.~\ref{equ1} and we will get the upper bound of $\lambda$:
\begin{equation}
{\lambda _{\max }} = \sqrt {{V_{x\max }}{V_{p\max }}}.\label{equ18}
\end{equation}

In fact, finding the confidence interval of $\lambda$ with confidence probability $\epsilon$ is a more precise method to estimate the upper bound of $\lambda$. But the calculation is too complex for the distribution of product of two Gaussian distributions which don't satisfy standard Gaussian distribution~\cite{springer1970distribution}. Fortunately, the value of $\lambda$ calculated by the upper bounds of $V_x$ and $V_p$ is very close to the more precise one~\cite{leverrier2010finite}. So, we can use $\lambda_{max}$ to estimate the extractable randomness.

\begin{figure}[t]
  \centering
  \includegraphics[width=.5\textwidth]{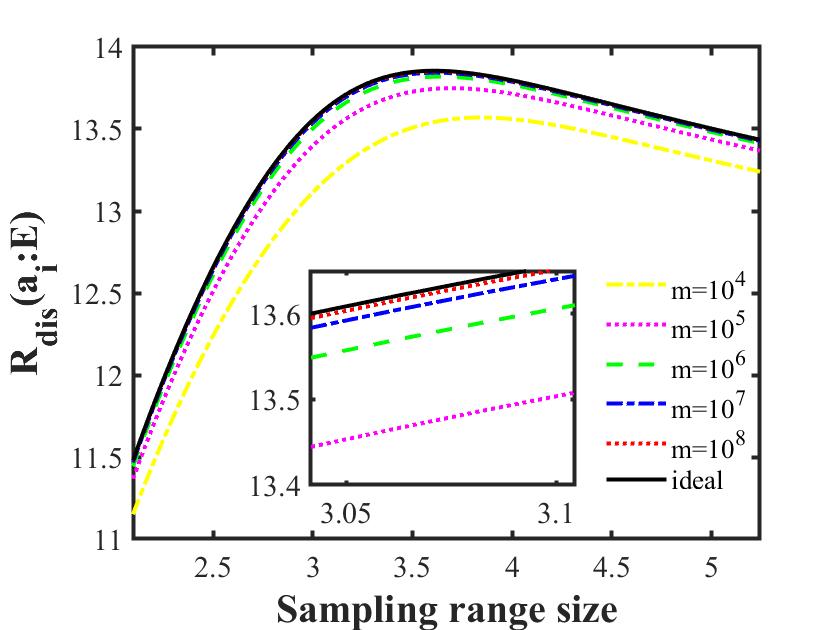}
  \caption{(Color online) The impact of sampling range size $N$ for finite-size effect on extractable randomness $R_{dis}(a_i:E)$ with different check data length $m$, which from bottom to top curve correspond to $m=10^4,10^5,10^6,10^7,10^8$. We set the excess noise $\varepsilon=0.1$, ideal variance $\sigma^2=1+\varepsilon=1.1$, sampling resolution $n=16$, and confidence probability $\epsilon=10^{-10}$ here.}
  \label{img4}
\end{figure}

\begin{figure}[t]
  \centering
  \includegraphics[width=.5\textwidth]{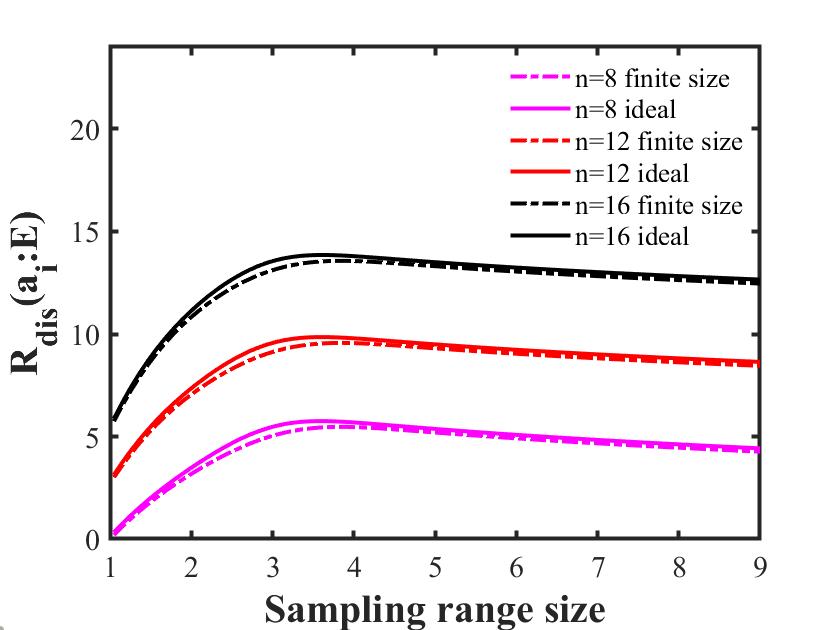}
  \caption{(Color online) The impact of sampling range size $N$ for finite-size effect on extractable randomness $R_{dis}(a_i:E)$ with different sampling resolution $n$, which we choose $n=8,12,16$. We set the excess noise $\varepsilon=0.1$, ideal variance $\sigma^2=1+\varepsilon=1.1$, check data length $m=10^4$, and confidence probability $\epsilon=10^{-10}$ here.}
  \label{img5}
\end{figure}

\section{simulation results and discussion }
\label{sec4}

In this section, we'll have some numerical simulations and discuss the results for considering finite-size on CV-SI-QRNG with the ideal measurement assumption. We assume that the input state is vacuum state here, i.~e.~the X and P quadratures satisfy the same probability distribution. It leads to ${V_{x\max }}{\rm{ = }}{V_{p\max }}$. The large bound $\left[ {{\rm{ - }}{a_{\lim }},{a_{\lim }}} \right]$, which is defined to determine the $a_{max}$ and $a_{min}$ in Eq.~\ref{equ6}, is set as $\left[ {{\rm{ - 10}}{\sigma},10{\sigma}} \right]$. The probability that $a_i$ falls out of the large bound is less than $1.5 \times {10^{{\rm{ - }}23}}$. \\

Fig.~\ref{img2} shows the relationship between extractable randomness and check data length. The red solid line is the ideal extractable randomness in infinite-size case, and the blue dot-dashed line is the extractable randomness with the effect of finite size. We can see that, the blue line changes sharply since check data length reduces under $10^{4}$ (some assumptions we made above have large deviation here), i.~e.~finite-size effect has remarkable influence on finial randomness. When check data length is larger than $10^{7}$, the finite-size line is very close to ideal line. In practical, we should choose an appropriate check data length to obtain enough final randomness meanwhile use less computing resources.

The another important parameter of finite-size effect is confidence probability $\epsilon$. The relationship between extractable randomness and confidence probability is shown in Fig. \ref{img3}. The curve of finite-size is more smoothly as $\epsilon$ decreases. At the point $\epsilon=1$, the finite-size effect on extractable randomness disappears due to the estimation for confidence interval of variance converges towards its ideal value. The confidence probability can be used as security parameter to adjust final randomness for different demands on security.

The function of sampling range size and extractable randomness under finite-size effect with different check data lengths has been displayed in Fig.~\ref{img4}. The curve is closer to the ideal curve (black solid line) with longer check data length. Finite-size effect is relatively smooth after the increase of sampling range size over the value that achieves max extractable randomness. In fact, as the increase of sampling range size, deviation between ideal and finite-size reduces very slowly. The gap between them is $0.1943$~bit when $N = 10 {\sigma}$, and reduces to $0.1924$~bit when $N = 100 {\sigma}$ with check data length $m = 10^4$, sampling resolution $n=16$, and confidence probability $\epsilon=10^{-10}$. We would like to point out that the most serious impact of finite-size appears in the peak value of extractable randomness, the same in Fig.~\ref{img5}, due to the comparatively uniform distribution of $a_i$. We discuss it in details in appendix A.

\begin{figure}[t]
  \centering
  \includegraphics[width=.5\textwidth]{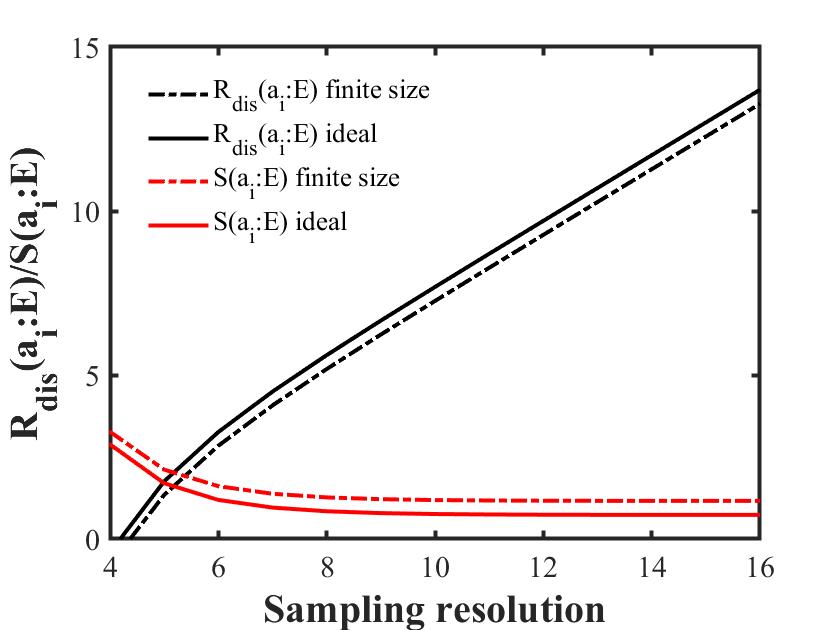}
  \caption{(Color online) Simulation results for upper bound of Eve's information $S(a_i:E)$ and extractable randomness $R_{dis}(a_i:E)$ as a function of sampling resolution $n$ under ideal infinite-size and finite-size condition. We set the excess noise $\varepsilon=0.1$, ideal variance $\sigma^2=1+\varepsilon=1.1$, sampling range size $N=3\sigma$, check data length $m=10^4$, and confidence probability $\epsilon=10^{-10}$ here.}
  \label{img6}
\end{figure}

In Fig.~\ref{img5}, we show the function of sampling range size and extractable randomness under finite-size effect with different sampling resolution. We choose $n=8, 12, 16$ three common sampling resolutions to show the influence. And we can see that, the finite-size effect has similar impact with different sampling resolutions. Fig.~\ref{img6} shows this impact in detail. We fix sampling range to $N =3{\sigma}$ and display the extractable randomness $R_{dis}(a_i:E)$ and the upper bound of Eve's information $S(a_i:E)$ as a function with sampling resolution $n$ under ideal infinite-size and finite-size conditions. The gap between the ideal curve and finite-size one is almost fixed with different sampling resolutions for both $R_{dis}(a_i:E)$ and $S(a_i:E)$.

\section{conclusion}
\label{sec5}
In this paper, we propose a method considering impact of finite-size effect on continuous variable source-independent quantum random number generation and improve the security of our protocol in practical applications. The method provides some formulas through central limit theorem for calculating the statistical fluctuations of parameters under finite-size scenario. Some numerical simulations are provided to show the influence on extractable randomness with several key variable parameters. We discuss these results in detail and summarize some conclusions, which could be useful in practical realization.

\begin{figure}[t]
  \centering
  \includegraphics[width=.5\textwidth]{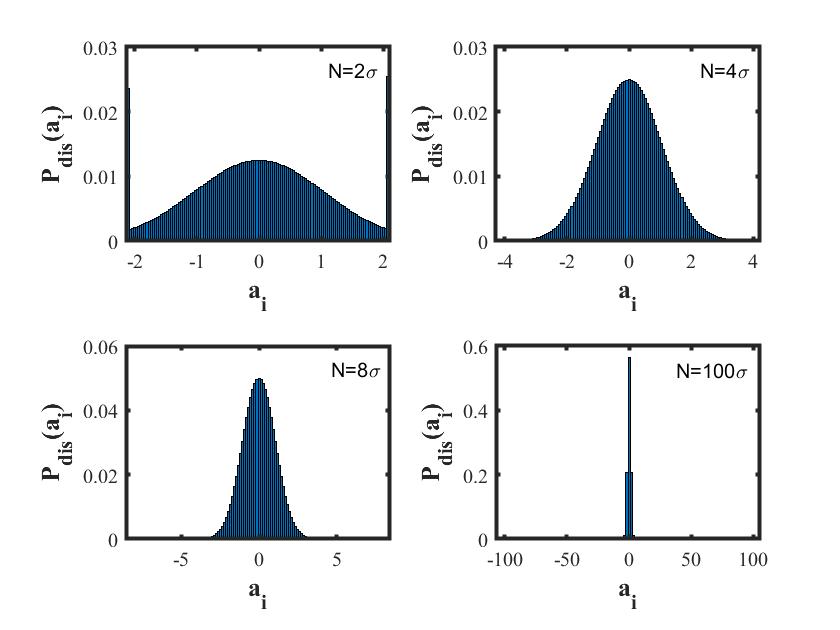}
  \caption{(Color online) The probability distribution $P_{dis}(a_i)$ of Alice's quantified output with different sampling range size $N=2\sigma,4\sigma,8\sigma,100\sigma$. We set the excess noise $\varepsilon=0.1$, ideal variance $\sigma^2=1+\varepsilon=1.1$ and fixed sampling resolution $n=8$ here.}
  \label{img7}
\end{figure}

The outcomes we found in numerical simulations show the rizations of finite-size effect. The final randomness is closer to the ideal value, with the increase of check data length and confidence probability, i.~e.~the effect of finite-size is monotonically related with these two parameters mightily. One could choose appropriate values according to the need for security and available computing resources in practical. An apposite sampling range size will reach the maximum of extractable randomness, which makes a relatively uniform distribution of input and the largest loss of randomness due to finite-size effect. Moreover, the loss is smoothly decreasing after sampling range size rises over the peak. While the impact of sampling resolution is more steady.

Our method still needs some improvements which can get further study, such as the treatment of the most border sampling value and the approximate on $\lambda_{max}$. In addiction,this work can promote the applications of continuous variable source-independent quantum random number generation and ensure the security in its processing.

Note added. Finite-size effect of the continuous variable source-independent quantum random number generation protocols has also been taken into consideration in~\cite{marangon2017source}, where the entropic uncertainty principle used for finite-size analysis through estimating the smooth min-conditional entropy. However, the central limit theorem and the impact of some key parameters haven't been considered for the finite-size analysis in detail.

\section*{Acknowledgements} \label{sec:acknowledgements}
This work was supported by the Key Program of National Natural Science Foundation of China under Grant No. 61531003, the National Natural Science Foundation under Grant No. 61427813, the Fund of CETC under Grant No. 6141B08231115, China Postdoctoral Science Foundation under Grant 2018M630116 and the Fund of State Key Laboratory of Information Photonics and Optical Communications.

\begin{figure}[t]
  \centering
  \includegraphics[width=.5\textwidth]{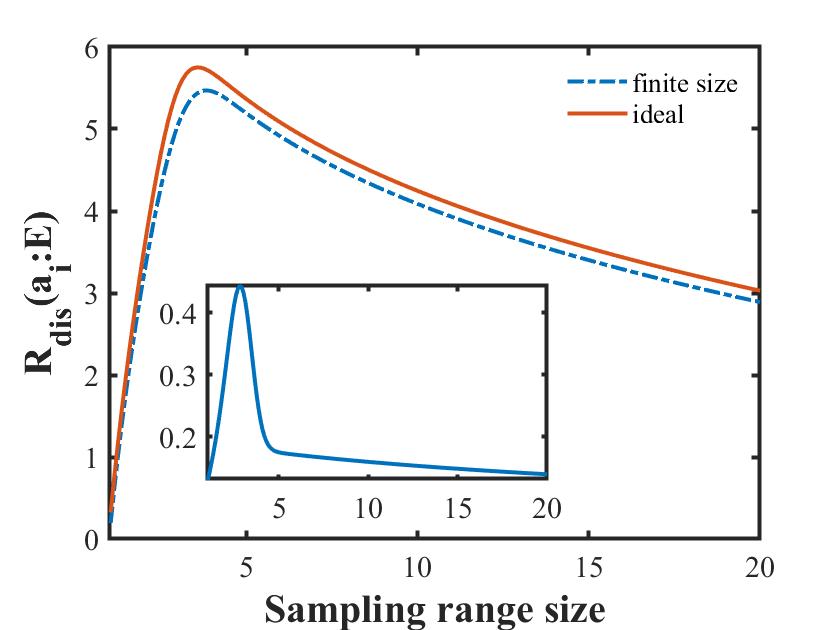}
  \caption{(Color online) Simulation result for extractable randomness $R_{dis}(a_i:E)$  as a function of sampling range size $N$ under ideal infinite-size and finite-size scenario. The mini figure inside is the difference between these two scenarios. We set the excess noise $\varepsilon=0.1$, ideal variance $\sigma^2=1+\varepsilon=1.1$, check data length $m=10^4$, confidence probability $\epsilon=10^{-10}$ and sampling resolution $n=8$ here.}
\label{img8}
\end{figure}

\appendix
\section{Effect of sampling range size}
\label{sec:appendix}

In this appendix, we focus on the impact of sampling range size on finite-size effect by analyzing the loss of extractable randomness, because of the remarkable influence of it for extractable randomness in both ideal and finite-size conditions.

This parameter mainly influences the probability distribution of Alice's discrete results as shown in Fig.~\ref{img7}. The probability will concentrate in the middle bins with the increase of the sampling range size. And if the sampling range size $N$ is too small, most of the probability will lie in boundary bins.

Fig.~\ref{img8} shows the extractable randomness $R_{dis}(a_i:E)$ with a larger sampling range size bound compared with Fig.~\ref{img4} and the difference value between ideal and finite-size conditions is shown in the inner mini figure. We can see that the curve of difference values has a peak when $N = 2.7\sigma$, and extractable randomness rises and falls sharply near the peak. But after the sampling range size increases over $5\sigma$, the reduction of difference values suddenly becomes very smooth.

The sampling range size value $N$ near peak is exactly the value that makes the most uniform probability distribution of $a_i$. When the sampling range size is small, compared to that with large sampling range size, the change of probability distribution is more conspicuous as the sampling range size changes. This corresponds the sharp and smooth changes in the mini figure of Fig.~\ref{img8} respectively.

It should be noted that the peak of extractable randomness on ideal infinite case appears in $N=3.4\sigma$ and doesn't coincide with the peak of difference. It leads to the max extractable randomness under finite-size condition appears in $N=3.7\sigma$, slightly larger than the value in ideal case. It should be paid attention in practical for obtaining maximal final randomness.

\bibliographystyle{apsrev4-1}
\bibliography{ref}

\end{document}